\documentclass[journal,twoside,9pt]{IEEEtran}

\usepackage{graphicx,epsfig,subfigure,amsmath,amssymb,enumerate,subeqnarray}
\usepackage{srcltx,floatflt,wrapfig,psfrag,kotex}

\usepackage[sort]{cite}
\usepackage{dsfont,
           amssymb,
            enumerate,
            color,
            mathrsfs,
            subfigure,
             psfrag,
             epsfig,
            }

\usepackage{multicol}
\usepackage{epsfig,psfrag}

\newtheorem{example}{Example}

\newtheorem{theorem}{Theorem}

\newtheorem{remark}{Remark}
\newtheorem{corollary}{Corollary}

\newtheorem{lemma}{Lemma}

\allowdisplaybreaks

\begin{document}

%
\title{Relaxed Conditions for Parameterized Linear Matrix Inequality in the Form of Double Sum}
%
%
%

\author{Do~Wan~Kim and Dong~Hwan~Lee,~\IEEEmembership{Member,~IEEE}
 \thanks{D. W. Kim is with the Department
of Electrical Engineering, Hanbat National University, Deajeon,
Korea, e-mail: dowankim@hanbat.ac.kr.}

\thanks{D. H. Lee is with the Department
of Electrical Engineering, KAIST, Deajeon,
Korea, e-mail: donghwan@kaist.ac.kr.}
}


\maketitle

\begin{abstract}
The aim of this study is to investigate less conservative conditions for a parameterized linear matrix inequality (PLMI) expressed in the form of a double convex sum. This type of PLMI frequently appears in T-S fuzzy control system analysis and design problems. In this letter, we derive new, less conservative linear matrix inequalities (LMIs) for the PLMI by employing the proposed sum relaxation method based on Young's inequality. The derived LMIs are proven to be less conservative than the existing conditions related to this topic in the literature. The proposed technique is applicable to various stability analysis and control design problems for T-S fuzzy systems, which are formulated as solving the PLMIs in the form of a double convex sum. Furthermore, examples is provided to illustrate the reduced conservatism of the derived LMIs.
\end{abstract}


\section{Introduction}

Parameterized Linear Matrix Inequalities (PLMIs) play a significant role in various fields, including T-S fuzzy control~\cite{Kong2020new,Tuan2001,Kim2000,Teixeira2003,Tanaka2004,guerra2006,guerra2012,Fang2006,Sala2007} and robust control~\cite{Oliveira1999,Peaucelle2000,Oliveira2002,Ramos2002,Leite2003,Oliveira2007,Nguyen2018}. These PLMIs are utilized in control design and analysis problems, where the objective is to find matrix variables that satisfy the PLMIs over infinite-dimensional parameter spaces. This problem, referred to as a PLMI problem (PLMIP) in this letter, poses challenges due to its infinite-dimensional nature, making it generally numerically intractable.

To address this challenge, researchers have made efforts to derive finite-dimensional sufficient linear matrix inequality (LMI) conditions for PLMIPs. For instance, several studies such as~\cite{Oliveira1999,Peaucelle2000,Oliveira2002} proposed sufficient LMI conditions for PLMIPs expressed as single convex sums (or fuzzy sums) of matrices. In these approaches, the matrices in the fuzzy sum vary within a matrix polytope as parameters change. Additionally, works such as~\cite{Tuan2001,Kim2000,Ramos2002,Teixeira2003,Leite2003,Tanaka2004,guerra2006,guerra2012} developed sufficient LMI conditions for PLMIPs expressed as double fuzzy sums. Moreover, techniques based on Polya's theorem~\cite{young1934} have been employed to extend these conditions to triple fuzzy sums~\cite{Fang2006} and more general $N$ fuzzy sums~\cite{Sala2007,Oliveira2007}. These sufficient conditions typically evaluate PLMIs at finite vertices of matrix polytopes, which cover the corresponding parameterized matrix space. The idea behind improving these conditions is to express matrix polytopes with finer and tighter vertex grids, and then verify the PLMIs at these vertices. This approach leads to less conservative conditions, reflecting the main concept of generalization based on Polya's theorem.

In this letter, our primary focus is on the sufficient LMI condition presented in~\cite{Tuan2001} for PLMIs expressed as a double fuzzy sum. This particular condition, as described in~\cite{Tuan2001}, incorporates distinct over-bounding techniques and employs different types of vertices compared to other existing works. It is well-known for producing less conservative results in numerous problem scenarios. The effectiveness of this sufficient condition has led to its widespread adoption within the field of fuzzy control, as shown in studies such as~\cite{guerra2012}. However, despite its initial development in 2001, to the best of our knowledge, its full potential has yet to be fully explored in subsequent research over the past few decades.

Motivated by the preceding discussion, the primary objective of this letter is to investigate less conservative LMI conditions for PLMIPs expressed as double fuzzy sums. This is achieved by extending the concepts introduced in~\cite{Tuan2001}. Specifically, we propose a novel sufficient LMI condition for PLMIPs using Young's inequality. The newly developed LMI condition encompasses the condition presented in~\cite{Tuan2001} as a special case. Theoretical analysis confirms that our proposed condition is less conservative than the one in~\cite{Tuan2001}. Additionally, our approach offers a clear and simplified analysis framework and proof for the condition in~\cite{Tuan2001} by leveraging Young's inequality, which can be generalized to PLMIPs with multiple fuzzy sum forms. The proposed technique is applicable to various control design problems in the context of T-S fuzzy systems represented by PLMIs formulated as double fuzzy sums. Furthermore, examples are included to demonstrate the decreased conservatism of the derived LMIs.

\textit{Notations:} The notation $P\succ Q$ ($P\prec Q$) denotes that the matrix $P-Q$ is positive (negative) definite. In this context, $\mathbb{R}$ represents the set of real numbers, $\mathbb{R}^{n}$ denotes the $n$-dimensional Euclidean space, and $\mathbb{R}^{m\times n}$ refers to the field of real matrices with dimensions $m\times n$. Furthermore, $\mathbb{I}_{r}$ represents the integer set $\{1,2,\ldots,r\}$ where $r>1$. To simplify notation, we will use $x$ instead of $x(t)$ for continuous-time signal vectors unless otherwise specified.

\section{Problem Formulation}

Consider the PLMI in the form of a double fuzzy sum:
\begin{align}
\Phi(z)=\sum_{i=1}^{r}\sum_{j=1}^{r}h_{i}(z)h_{j}(z)\Phi_{ij}\prec0
\label{eq: PLMI}
\end{align}
Here, $z\in\mathbb{R}^{p}$ represents the premise variable, $h_{i}(z):\mathscr{D}_{z}~\rightarrow \mathbb{R}_{[0,1]}$ satisfies $\sum_{i=1}^{r}h_{i}(z)=1$, $\mathscr{D}_{z}\subset \mathbb{R}^{p}$ denotes the domain of the premise variables, and $\Phi_{ij}=\Phi_{ij}^{T}\in\mathbb{R}^{n\times n},$ $(i,j)\in\mathbb{I}_{r}\times\mathbb{I}_{r}$ is a matrix that may have a linear dependency on the decision variables. The PLMI~\eqref{eq: PLMI} commonly arises in controller designs or stability analyses of linear uncertain models~\cite{Nguyen2018} and T-S fuzzy systems~\cite{Tuan2001,Tanaka2004,Kim2010}. The sufficient LMIs for negativeness, $\Phi_{ij}\prec0$, $(i,j)\in\mathbb{I}_{r}\times\mathbb{I}_{r}$, can be derived straightforwardly by omitting $h_{i}h_{j}\in\mathbb{R}_{>0}$. Various efforts have been made to provide less conservative LMI conditions for the PLMI~\eqref{eq: PLMI} in the form of a double fuzzy sum. Initially, relaxation techniques~\cite{Tuan2001,Tanaka2004} were introduced, utilizing the properties of double fuzzy sums, such as $\sum_{i=1}^{r}\sum_{j=1}^{r}h_{i}h_{j}\Phi_{ij}=\sum_{i=1}^{r}h_{i}^{2}\Phi_{ii}+\sum_{i=1}^{r}\sum_{j=1,j>i}^{r}h_{i}h_{j}(\Phi_{ij}+\Phi_{ji})$. These sum relaxations~\cite{Tuan2001,Tanaka2004} serve as the foundation for subsequent researches, including the introduction of slack variables~\cite{Kim2000,Teixeira2003} and the extension of multidimensional summations~\cite{Sala2007} (for more details, see~\cite{Sugeno2019}). As stated in~\cite{guerra2006,guerra2012}, the following lemma~\cite{Tuan2001} is one of popular LMI conditions:
\begin{lemma}[Theorem~2.2 of \cite{Tuan2001}]\label{lemma: relaxation for dubled sum}
The PLMI~\eqref{eq: PLMI} holds if
\begin{align*}
    \Phi_{ii} &\prec 0,\quad \forall i\in\mathbb{I}_{r}\\
    \frac{2}{r-1}\Phi_{ii}+\Phi_{ij}+\Phi_{ji}&\prec 0,\quad \forall
    (i,j)\in\{(i,j)\in\mathbb{I}_{r}\times\mathbb{I}_{r}~|~ i\neq j\}.
\end{align*}
\end{lemma}

Lemma~\ref{lemma: relaxation for dubled sum} will play a crucial role in this letter, as we develop a new LMI condition that includes it as a special case. In particular, the problems of interest in this letter is to develop less conservative LMI conditions which generalize those in Lemma~\ref{lemma: relaxation for dubled sum} and prove their less conservatism.

\section{Main Results}

Before presenting our main results, we introduce the following lemmas, which will be used throughout the letter:
\begin{lemma}[Young's inequality \cite{young1934}]\label{lemma: young}
For  $a\in\mathbb{R}_{\geqslant 0}$ and $b\in\mathbb{R}_{\geqslant 0}$, $ab\le \frac{1}{2}{a^2} + \frac{1}{2}{b^2}$ holds.
\end{lemma}

\begin{lemma}\label{lemma: Xij}
For any matrix $\Xi_{ij}$, $(i,j)\in\left\{\left. (i,j)\in\mathbb{I}_{r}\times \mathbb{I}_{r}~\right|~ j\neq i \right\}$,
it is true that
\begin{align*}
    \sum_{i=1}^{r}
    \sum_{j=1,j\neq i}^{r}
    h_{i}^{2} (\Xi_{ij}+\Xi_{ji})
    =
    \sum_{i=1}^{r}
    \sum_{j=1,j\neq i}^{r}
    h_{j}^{2} (\Xi_{ij}+\Xi_{ji}).
\end{align*}
\end{lemma}
\begin{IEEEproof}
See Appendix~\ref{app: lemma Xij}
\end{IEEEproof}

Now, we are ready to present our main result. In the following theorem, we present a new relaxed LMI condition using Young's inequality, that generalizes Lemma~\ref{lemma: relaxation for dubled sum}.
\begin{theorem}\label{th: main}
The PLMI \eqref{eq: PLMI} holds if
\begin{align}
\Phi _{ii}  + \frac{1}{2}\sum\limits_{j = 1,j < i}^r {\delta _j } (\Phi _{ij}  + \Phi _{ji} ) + \frac{1}{2}\sum\limits_{j = 1,j > i}^r {\delta _{j - 1} } (\Phi _{ij}  + \Phi _{ji} ) \prec 0
\label{eq: propose LMI}
\end{align}
for all $i \in {\mathbb I}_r$ and all
\[
(\delta _1 ,\delta _2 , \ldots ,\delta _{r - 1} ) \in \underbrace {\{ 0,1\}  \times \{ 0,1\}  \times  \cdots  \times \{ 0,1\} }_{(r - 1) - {\rm{times}}}.
\]
\end{theorem}
\begin{IEEEproof}
It is obvious that $\textrm{PLMI \eqref{eq: PLMI}} \Leftrightarrow x^{T}\Phi(z)x<0,\forall x \in\mathbb{R}^{n}, x \neq 0$. We can equivalently rewrite $x^{T}\Phi(z)x$ as
\begin{align*}
    x^{T} \Phi(z)x
    &=
    \sum_{i=1}^{r} {h_{i}^{2} x^{T} \Phi_{ii}x}
    +
    \frac{1}{2}
    \sum_{i=1}^{r}
    \sum_{j=1,j\neq i}^{r}
    h_{i} h_{j} x^{T} (\Phi_{ij}+\Phi_{ji})x
\end{align*}
If  $ x^{T}(\Phi_{ij}+\Phi_{ji}) x \geq 0$ on the right-hand side in the above equation, then since $ h_i h_j x^{T}(\Phi_{ij}+\Phi_{ji}) x \geq 0$, applying Lemma~\ref{lemma: young} to
$h_{i} h_{j} x^{T} (\Phi_{ij}+\Phi_{ji})x$ yields
\begin{align*}
    h_{i} h_{j} x^{T} (\Phi_{ij}+\Phi_{ji})x
    \leq
    \frac{1}{2}
    \left(
    h_{i}^{2}+h_{j}^{2}
    \right)x^{T} (\Phi_{ij}+\Phi_{ji})x
\end{align*}
Otherwise, if $ x^{T}(\Phi_{ij}+\Phi_{ji}) x <0$, then $h_{i} h_{j} x^{T} (\Phi_{ij}+\Phi_{ji})x\le 0$. By considering both cases, we have
\begin{align*}
    h_{i} h_{j} x^{T} (\Phi_{ij}+\Phi_{ji})x
    &\le
    \max
    \left\{
    \frac{1}{2}
    \left(
    h_{i}^{2}+h_{j}^{2}
    \right)x^{T} (\Phi_{ij}+\Phi_{ji})x,0
    \right\}
    \\
    &=
    \frac{1}{2}
    \left(
    h_{i}^{2}+h_{j}^{2}
    \right)
    \max
    \left\{
    x^{T} (\Phi_{ij}+\Phi_{ji})x,0
    \right\}.
\end{align*}
With this inequality, it can be shown that
\begin{align*}
    x^{T} \Phi(z)x
    \le &
    \sum_{i=1}^{r} {h_{i}^{2} x^{T} \Phi_{ii}x}
    \\
    &+
    \frac{1}{4}
    \sum_{i=1}^{r}
    \sum_{j=1,j \neq i}^{r}
    \left(
    h_{i}^{2}+h_{j}^{2}
    \right)
    \max
    \left\{
    x^{T} (\Phi_{ij}+\Phi_{ji})x,0
    \right\}.
\end{align*}
Using Lemma~\ref{lemma: Xij}, we obtain
\begin{align*}
    &
    x^{T} \Phi(z)x
    \\
    &\quad\le
    \sum_{i=1}^{r}{h_{i}^{2} x^{T} \Phi_{ii}x}
    +
    \frac{1}{2}
    \sum_{i=1}^{r}
    \sum_{j=1,j \neq i}^{r}
    h_{i}^{2}
    \max
    \left\{
    x^{T} (\Phi_{ij}+\Phi_{ji})x,0
    \right\}
    \\
    &\quad=
    \sum_{i=1}^{r}
    h_{i}^{2}
    \left(
    x^{T}\Phi_{ii}x
    +
    \frac{1}{2}
    \sum_{j=1,j\neq i}^{r}
    \max
    \left\{
    x^{T} (\Phi_{ij}+\Phi_{ji})x,0
    \right\}
    \right).
\end{align*}
To avoid $\max$ operator on the right-hand side, we can consider all possible cases, which leads to the desired conclusion:
\begin{align*}
\textrm{LMIs \eqref{eq: propose LMI}} \Rightarrow
     x^{T}\Phi(z)x<0,\forall x \in {\mathbb R}^n, x\neq 0 \Leftrightarrow \textrm{PLMI \eqref{eq: PLMI} }
\end{align*}
\end{IEEEproof}

Lemma~\ref{lemma: relaxation for dubled sum} is based on the positive definiteness property of a 2-by-2 matrix within the nonnegative orthant. In contrast, the proposed sufficient condition employs a different bounding approach utilizing Young's inequality.
Consequently, conducting a direct and intuitive comparison between the two may not be feasible. However, we can demonstrate that the proposed Theorem~\ref{th: main} is less conservative than Lemma~\ref{lemma: relaxation for dubled sum}. The utilization of Young's inequality, which provides more flexibility in calculating the new bounds, is the main factor contributing to the relaxation.
\begin{corollary}
\label{th:conservativeness}
The LMI condition~\eqref{eq: propose LMI} of Theorem~\ref{th: main} is less conservative than the LMI condition of Lemma~\ref{lemma: relaxation for dubled sum}.
\end{corollary}
\begin{IEEEproof}
To prove the claim about conservatism, we show that the LMIs of Lemma~\ref{lemma: relaxation for dubled sum} imply LMIs \eqref{eq: propose LMI} of Theorem~\ref{th: main}, while the converse is not true. First, one can prove that the LMIs of Lemma~\ref{lemma: relaxation for dubled sum} implies
\[
\frac{2}{{r - 1}}\Phi _{ii}  + \delta _k (\Phi _{ij}  + \Phi _{ji} ) \prec 0,\quad \forall i \in {\mathbb I}_r ,k \in \{ 1,2, \ldots ,r - 1\}
\]
for all $\delta_k\in \{ 0,1 \},k \in \{ 1,2, \ldots ,r - 1\}$. This is because when $\delta_k = 1$, the above LMI becomes the second LMI in Lemma~\ref{lemma: relaxation for dubled sum}, and when $\delta_k= 0$, the above LMI becomes the first LMI in Lemma~\ref{lemma: relaxation for dubled sum}. Next, letting
\[
k = k(j) = \left\{ {\begin{array}{*{20}c}
   j, & j < i;  \\
   j - 1, & {\rm{otherwise}}\\
\end{array}} \right.
\]
in the foregoing inequality and summing both sides of the foregoing inequality over $j\in \mathbb{I}_{r}, j\neq i$ lead to~\eqref{eq: propose LMI}. Therefore, this proves that if the LMIs of  of Lemma~\ref{lemma: relaxation for dubled sum} are feasible, then so are the LMIs in~\eqref{eq: propose LMI}.

Next, we provide an example for which the LMIs in~\eqref{eq: propose LMI} are feasible, while the LMIs of Lemma~\ref{lemma: relaxation for dubled sum} are not feasible. Consider the case that $r=3$, $\Phi_{11}=-2$, $\Phi_{22}= -1$, $\Phi_{33}=-2$, $\Phi_{12}= 0$, $\Phi_{13}=2$, $\Phi_{21}= 0$, $\Phi_{23}=-1$, $\Phi_{31}=  0$, and $\Phi_{32}= 0$.
Then, LMIs \eqref{eq: propose LMI} of Theorem~\ref{th: main} are satisfied as follows: $\Phi _{11}  + \sum_{j = 1,j < 1}^3 {\delta _j } \frac{{\Phi _{1j}  + \Phi _{j1} }}{2} + \sum_{j = 1,j > 1}^3 {\delta_{j-1} } \frac{{\Phi _{1j}  + \Phi _{j1} }}{2}$ is $\Phi_{11} =-2$ when $(\delta _1 ,\delta _2) = (0,0)$, $\Phi_{11}+ \frac{\Phi_{13}+\Phi_{31}}{2} =-0.5$ when $(\delta_1,\delta_2) = (0,1)$, $\Phi_{11}+ \frac{\Phi_{12}+\Phi_{21}}{2} = -2$  when $(\delta _1,\delta _2) = (1,0)$, $\Phi_{11} +\sum_{j=1,j\neq 1}^{3} \frac{\Phi_{1j}+\Phi_{j1}}{2} =-0.5$ when $(\delta _1,\delta _2) = (1,1)$. For $\Phi _{22}  + \sum_{j = 1,j < 2}^3 {\delta _j } \frac{{\Phi _{2j}  + \Phi _{j2} }}{2} + \sum_{j = 1,j > 2}^3 {\delta_{j-1} } \frac{{\Phi _{2j}  + \Phi _{j2} }}{2}$, it is equal to $\Phi_{22}=-1$ when $(\delta_1,\delta _2)=(0,0)$, $\Phi_{22} + \frac{\Phi_{23}+\Phi_{32}}{2}  =-1.5$ when $(\delta_1,\delta_2) = (0,1)$, $\Phi_{22} +\frac{\Phi_{21}+\Phi_{12}}{2} = -1$ when $(\delta_1,\delta_2) = (1,0)$, $\Phi_{22} +\sum_{j=1,j\neq 2}^{3} \frac{\Phi_{2j}+\Phi_{j2}}{2} = -1.5$ when $(\delta_1,\delta _2) = (1,1)$. Lastly, for $\Phi _{33}  + \sum_{j = 1,j < 3}^3 {\delta _j } \frac{{\Phi _{3j}  + \Phi _{j3} }}{2} + \sum_{j = 1,j > 3}^3 {\delta _{j - 1} } \frac{{\Phi _{3j}  + \Phi _{j3} }}{2}
$, it is equal to $\Phi_{33}=-2$ when $(\delta_1,\delta_2) = (0,0)$, $\Phi_{33} +\frac{\Phi_{32}+\Phi_{23}}{2} =-2.5$ when $(\delta_1,\delta _2) = (0,1)$, $\Phi_{33}+\frac{\Phi_{31}+\Phi_{13}}{2}=-0.5$ when $(\delta_1,\delta _2) = (1,0)$, $\Phi_{33} + \sum_{j=1,j\neq 3}^{3}\frac{\Phi_{3j}+\Phi_{j3}}{2} =-1$  when $(\delta_1,\delta _2) = (1,1)$.
On the other hand, the LMIs of Lemma~\ref{lemma: relaxation for dubled sum} are infeasible because $\frac{2}{{r - 1}}\Phi _{11}  + \Phi _{13}  + \Phi _{31}  =  - 2 + 2 + 0 = 0$. This completes the proof.

\end{IEEEproof}

Corollary~\ref{th:conservativeness} establishes that Theorem~\ref{th: main} is not more conservative than Lemma~\ref{lemma: relaxation for dubled sum} by demonstrating that the condition in Lemma~\ref{lemma: relaxation for dubled sum} implies the condition in Theorem~\ref{th: main}. Furthermore, it proves the less conservativeness of Theorem~\ref{th: main} compared to Lemma~\ref{lemma: relaxation for dubled sum} by presenting a counterexample for which the condition of Theorem~\ref{th: main} is feasible while the condition of Lemma~\ref{lemma: relaxation for dubled sum} is not.

\begin{remark}
We emphasize that
\begin{enumerate}
    \item When $r=2$, LMIs \eqref{eq: propose LMI} of Theorem \ref{th: main} are equal to those of Lemma \ref{lemma: relaxation for dubled sum}.
    \item Both Lemma \ref{lemma: relaxation for dubled sum} and Theorem \ref{th: main} need no additional slack variables.
    \item Lemma \ref{lemma: relaxation for dubled sum} still plays an important role when sufficient LMIs are derived from PLMI \eqref{eq: PLMI} in the form of the double convex sum (see, for example, \cite{Tapia2017,Elias2021,Li2021}, and references therein). However, Theorem \ref{th:conservativeness} shows that LMIs of Lemma \ref{lemma: relaxation for dubled sum} are sufficient to ensure that LMIs \eqref{eq: propose LMI} of Theorem \ref{th: main} hold; hence, Theorem \ref{th: main} could be substituted for Lemma \ref{lemma: relaxation for dubled sum}.
\end{enumerate}

\end{remark}

\begin{remark}
The numerical complexity of optimization problems based on LMIs can be estimated by considering the total number of scalar decision variables, denoted as $N_D$, and the total row size of the LMIs, denoted as $N_L$~\cite{gahinet1996lmi}. For the LMIs in Lemma~\ref{lemma: relaxation for dubled sum} and Theorem~\ref{th: main}, the number of scalar decision variables depends on the specific problem at hand. Therefore, we focus solely on the total row size $N_L$ of the LMIs.
In Lemma~\ref{lemma: relaxation for dubled sum}, we have $N_L = n \times r^{2}$, whereas in Theorem~\ref{th: main}, we have $N_L = n \times r \times 2^{r-1}$. It is worth noting that the total row size grows exponentially as $r$ increases. Hence, the numerical complexity of Theorem~\ref{th: main} is greater than that of Lemma~\ref{lemma: relaxation for dubled sum}, particularly when $r$ is large. This increased complexity can be regarded as a cost to pay for the relaxation.
\end{remark}

\begin{remark}
Lemma~\ref{lemma: relaxation for dubled sum} relies on the positive definiteness property of a 2-by-2 matrix within the nonnegative orthant. Consequently, generalizing this condition to triple and more general $N$ fuzzy summation cases is challenging due to the need for considering the positive definiteness property of general matrices, which are inherently more complex.
On the other hand, the proposed approach is based on Young's inequality, enabling a more straightforward generalization to cases involving more general $N$ fuzzy summations. Although the general $N$ fuzzy summations encompass a broader range of applications, it is worth noting that the simple double fuzzy summation case still encapsulates the core ideas. Considering these generalizations is deferred to potential future topics.
\end{remark}

\section{Examples}

All numerical examples in this letter were treated with the
help of MATLAB 2020a running on a Windows 10 PC with AMD Ryzen 3970X 3.69G Hz CPU, 128 GB RAM. The LMI problems were solved with MATLAB LMI Control Toolbox~\cite{gahinet1996lmi}.

\begin{example}\label{ex:1}
Consider the asymptotic stabilization condition~\cite{Tanaka2004} for continuous-time T-S fuzzy systems in the form of PLMI \eqref{eq: PLMI} with
\begin{align*}
    \Phi_{ij}
    =
    (A_{i}Q+B_{i}F_{j})^{T}+A_{i}Q+B_{i}F_{j}
    ,\quad (i,j)\in\mathbb{I}_{3}\times \mathbb{I}_{3}
\end{align*}
widely used for the fuzzy control system
\begin{align*}
\dot{x}=\sum_{i=1}^{3}h_{i}(A_{i}x+B_{i}u), \quad u=\sum_{i=1}^{3}h_{i}F_{i}Q^{-1}x,
\end{align*}
where $Q=Q^{T}\succ0$ and $F_{i}$ are the matrix variables to be determined, $A_{i}$ and $B_{i}$ borrowed from \cite{Fang2006} are
\begin{align*}
A_{1}=&\begin{bmatrix}
    1.59&-7.29\\0.01&0
    \end{bmatrix}, B_{1}=\begin{bmatrix}
    1\\0\end{bmatrix},\\
A_{2}=&\begin{bmatrix}
    0.02&-4.64\\0.35&0.21
    \end{bmatrix}, B_{2}=\begin{bmatrix}
    8\\0
    \end{bmatrix},\\
A_{3}=& \begin{bmatrix}
    -a&-4.33\\0&0.
    \end{bmatrix}, B_{3} =\begin{bmatrix}
    -b+6\\-1
    \end{bmatrix}.
\end{align*}
We change the values of the nonnegative parameters $a$ and $b$ to check the feasibility of LMIs of Lemma~\ref{lemma: relaxation for dubled sum} and Theorem~\ref{th: main}. Figure~\ref{fig:1} shows the feasible points $(a,b)$ for Lemma~\ref{lemma: relaxation for dubled sum} and Theorem~\ref{th: main}, respectively. Here, the mark `$\circ$' indicates that the LMI of Theorem~\ref{th: main} is feasible, and the mark `$\times$' denotes that the LMI of Lemma~\ref{lemma: relaxation for dubled sum} is feasible. As discussed in Theorem~\ref{th:conservativeness}, one can see that Theorem~\ref{th: main} provides less conservative results than Lemma~\ref{lemma: relaxation for dubled sum}.

\begin{figure}\label{fig:1}
\centering
\includegraphics[width=0.5\textwidth]{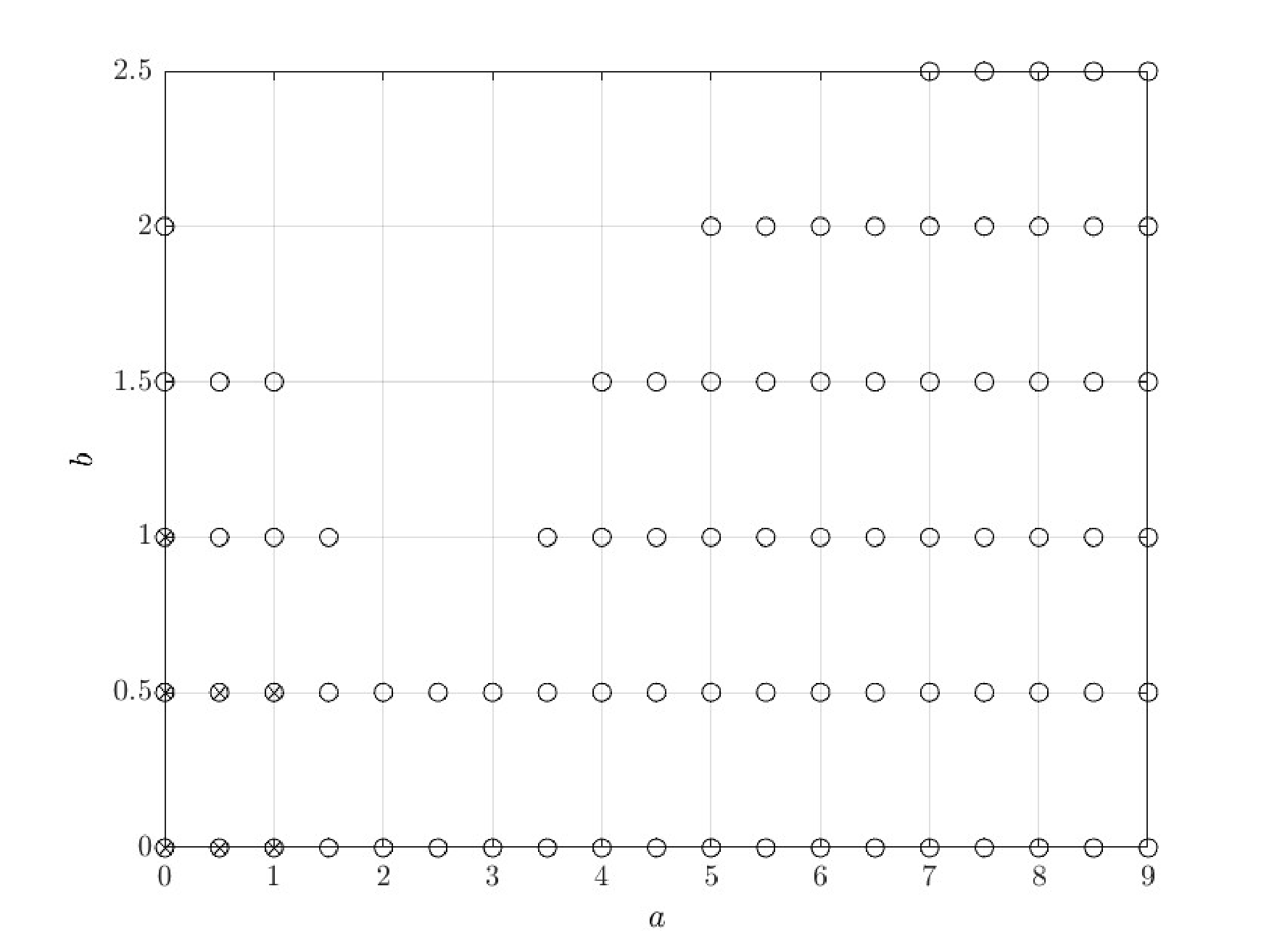}
\caption{Example~\label{ex:1}: Stabilizable region by Theorem~\ref{th: main} (`$\circ$') and Lemma~\ref{lemma: relaxation for dubled sum} (`$\times$') }
\end{figure}
\end{example}
\begin{example}\label{ex:2}
Consider the discrete-time T-S fuzzy control system
\begin{align*}
x(k+1) =&\sum_{i=1}^{3}h_{i}(A_{i}x(k)+B_{i}u(k)), \\
u(k) =&\sum_{i=1}^{3}h_{i}F_{i}Q^{-1}x(k),
\end{align*}
where $Q=Q^{T}\succ0$ and $F_{i}$ are the matrix variables to be determined, $A_{i}=I+\tilde{A}_{i}T$, $B_{i}=\tilde{B}_{i}T$ for $i\in\mathbb{I}_{3}$, $T=0.4$, and
\begin{align*}
\tilde{A}_{1}=&
    \begin{bmatrix}
    2&-10\\1&0
    \end{bmatrix}, \tilde{B}_{1}
    = \begin{bmatrix}
    1\\0
    \end{bmatrix},\\
\tilde{A}_{2}=&
    \begin{bmatrix}
    a&-10\\1&0
    \end{bmatrix}, \tilde{B}_{2}
    =
    \begin{bmatrix}
    b\\0
    \end{bmatrix},\\
 \tilde{A}_{3}=&     \begin{bmatrix}
    -2&-10\\1&0
    \end{bmatrix}, \tilde{B}_{3}
    =
    \begin{bmatrix}
    1\\0.334
    \end{bmatrix}
\end{align*}
borrowed from~\cite{Sala2007b}. The asymptotic stabilization condition~\cite{Tanaka2004} takes the PLMI form~\eqref{eq: PLMI} with
\begin{align*}
\Phi_{ij}=\begin{bmatrix}
        -Q&(A_{i}Q+B_{i}F_{j})^{T}\\
        A_{i}Q+B_{i}F_{j}&-Q
    \end{bmatrix}.
\end{align*}
We find the feasible regions of Lemma~\ref{lemma: young} and Theorem~\ref{th: main} by varying the values of nonnegative $a$ and $b$. Figure~\ref{fig:2} depicts the feasible points $(a,b)$,
where the mark `$\times$' and `$\circ$' are for Lemma~\ref{lemma: young} and Theorem~\ref{th: main}, respectively.
As shown Figure~\ref{fig:2}, we see that the feasible region of Theorem~\ref{th: main} is larger than that of Lemma~\ref{lemma: young}.
\begin{figure}\label{fig:2}
\centering
\includegraphics[width=0.5\textwidth]{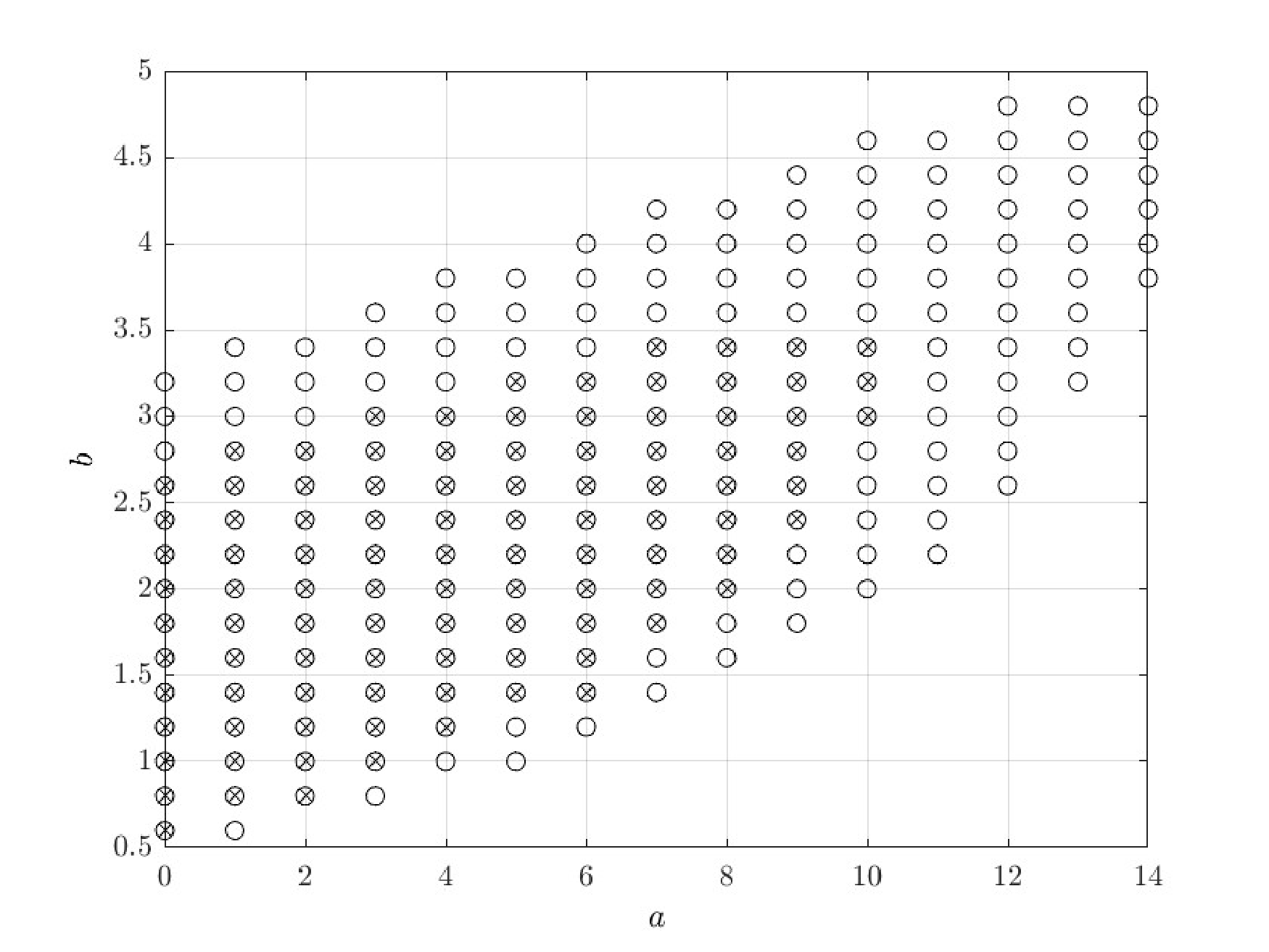}
\caption{Example~\label{ex:2} Stabilizable region by Theorem~\ref{th: main} (`$\circ$') and Lemma~\ref{lemma: relaxation for dubled sum} (`$\times$') }
\end{figure}

\end{example}

\section{Conclusions}

This letter presents new sufficient conditions for PLMIs formulated as a double convex sum. These conditions are expressed in LMIs, derived without the need for slack variables. Moreover, the less conservative nature of these conditions has been rigorously demonstrated. Theoretical claims put forth in this work have been effectively validated through experimental results.
Potential avenues for future research include generalizing the LMI conditions for general $N$ convex summations. Additionally, conducting a comparative analysis between the proposed frameworks and slack variable approaches documented in the literature presents an interesting prospect for further investigation.

Finally, it is worth noting that although this letter specifically focuses on double fuzzy sum cases, the concept can be readily extended to encompass triple and more general $N$ fuzzy sum cases, albeit with increased complexity.
However, expanding the analysis to these scenarios significantly complicates the main analysis, which exceeds the scope of this letter. Moreover, it has the potential to obscure the fundamental ideas and insights of our proposed approach.
Hence, this letter restricts its scope to the double fuzzy sum scenarios, and we defer the exploration of these extensions to future research, as they hold promising possibilities for further investigation.

\appendices
\section{Proof of Lemma \ref{lemma: Xij} \label{app: lemma Xij}}
It is not hard to see that

\begin{align*}
    &
    \sum_{i=1}^{r}
    \sum_{j=1,j\neq i}^{r}
    h_{i}^{2} (\Phi_{ij}+\Phi_{ji})
    \\
    &=
    \sum_{i=1}^{r}
    \sum_{j=1,j\neq i}^{r}
    (h_{i}^{2}-h_{j}^{2}) (\Phi_{ij}+\Phi_{ji})
    +
    \sum_{i=1}^{r}
    \sum_{j=1,j\neq i}^{r}
    h_{j}^{2} (\Phi_{ij}+\Phi_{ji})
    \\
    &=
    \sum_{i=1}^{r}
    \sum_{j=1,j>i}^{r}
    (h_{i}^{2}-h_{j}^{2}) (\Phi_{ij}+\Phi_{ji})
    \\
    &\quad
    +
    \sum_{i=1}^{r}
    \sum_{j=1,j<i}^{r}
    (h_{i}^{2}-h_{j}^{2}) (\Phi_{ij}+\Phi_{ji})
    +
    \sum_{i=1}^{r}
    \sum_{j=1,j\neq i}^{r}
    h_{j}^{2} (\Phi_{ij}+\Phi_{ji})
    \\
    &=
    \sum_{i=1}^{r}
    \sum_{j=1,j>i}^{r}
    (h_{i}^{2}-h_{j}^{2}) (\Phi_{ij}+\Phi_{ji})
    \\
    &\quad
    -
    \sum_{i=1}^{r}
    \sum_{j=1,j>i}^{r}
    (h_{i}^{2}-h_{j}^{2}) (\Phi_{ij}+\Phi_{ji})
    +
    \sum_{i=1}^{r}
    \sum_{j=1,j\neq i}^{r}
    h_{j}^{2} (\Phi_{ij}+\Phi_{ji})
    \\
    &=
    \sum_{i=1}^{r}
    \sum_{j=1,j\neq i}^{r}
    h_{j}^{2} (\Phi_{ij}+\Phi_{ji}).
\end{align*}

\bibliographystyle{IEEEtran}
\bibliography{reference}

\end{document}